\documentclass[preprint2,longabstract]{aastex}
\usepackage{natbib}
\usepackage{color}
\shorttitle{Dynamical structure of IRAS 16293-2422}
\shortauthors{Favre et al.}

\begin{document}

\title{Dynamical structure of the inner 100~AU of the deeply embedded protostar IRAS~16293-2422}

\author{C\'ecile Favre}
\affil{Department of Physics and Astronomy, University of Aarhus, Ny Munkegade 120, 8000 Aarhus C, Denmark}
\affil{Department of Astronomy, University of Michigan, 500 Church St., 
    Ann Arbor, MI 48109, USA}
\email{cfavre@umich.edu}

\and

\author{Jes~K. J\o rgensen}
\affil{Centre for Star and Planet Formation, Niels Bohr Institute, University of Copenhagen, Juliane Maries Vej 30, 2100 Copenhagen \O, Denmark}
\affil{Natural History Museum of Denmark, University of Copenhagen, {\O}ster Voldgade 5-7, 1350 Copenhagen K., Denmark}
\and

\author{David Field }
\affil{Department of Physics and Astronomy, University of Aarhus, Ny Munkegade 120, 8000 Aarhus C, Denmark}

\and

\author{Christian Brinch, Suzanne~E. Bisschop}
\affil{Centre for Star and Planet Formation, Niels Bohr Institute, University of Copenhagen, Juliane Maries Vej 30, 2100 Copenhagen \O, Denmark}
\affil{Natural History Museum of Denmark, University of Copenhagen, {\O}ster Voldgade 5-7, 1350 Copenhagen K., Denmark}

\and

\author{Tyler~L. Bourke}
\affil{Harvard-Smithsonian Center for Astrophysics, 60 Garden Street MS42, Cambridge, MA 02138, USA}
\affil{SKA Organisation, Jodrell Bank Observatory, Lower Withington, Macclesfield, Cheshire, SK11 9DL, UK}

\and

\author{Michiel~R. Hogerheijde}
\affil{Leiden Observatory, Leiden University, P.O. Box 9513, 2300 RA Leiden, The Netherlands}

\and

\author{Wilfred~W.~F. Frieswijk}
\affil{Netherlands Institute for Radio Astronomy, Postbus 2, 7990 AA, Dwingeloo, The Netherlands}

%
\begin{abstract}

A fundamental question about the early evolution of low-mass protostars is when circumstellar disks may form. High angular resolution observations of molecular transitions in the (sub)millimeter wavelength windows make it possible to investigate the kinematics of the gas around newly-formed stars, for example to identify the presence of rotation and infall. IRAS16293-2422 was observed with the extended Submillimeter Array (eSMA) resulting in subarcsecond resolution (0.46$\arcsec$ $\times$ 0.29$\arcsec$, i.e. $\sim$ 55 $\times$ 35~AU) images of compact emission from the C$^{17}$O (3-2) and C$^{34}$S (7-6) transitions at 337~GHz (0.89~mm). To recover the more extended emission we have combined the eSMA data with SMA observations of the same molecules. The emission of C$^{17}$O (3-2) and C$^{34}$S (7-6) both show a velocity gradient oriented along a northeast-southwest direction with respect to the continuum marking the location of one of the components of the binary, IRAS16293A. 
Our combined eSMA and SMA observations show that the velocity field on the 50--400~AU scales is consistent with a rotating structure. It cannot be explained by simple Keplerian rotation around a single point mass but rather needs to take into account the enclosed envelope mass at the radii where the observed lines are excited. We suggest that IRAS~16293-2422 could be among the best candidates to observe a pseudo-disk with future high angular resolution observations.
\end{abstract}
 
\keywords{ISM: individual objects: IRAS 16293-2422 --- Radio lines: ISM --- Stars: circumstellar matter --- Stars: formation}

%
\section{Introduction}\label{sec:Introduction}
%
\begin{table*}
\begin{center}
\caption{Main parameters of eSMA and SMA datasets.\label{tab1}}
\begin{tabular}{ccccccccc}
\tableline\tableline
Set\tablenotemark{a} & Array &Observed&Rest & Spectral & \multicolumn{2}{c}{Synthesized}   & Flux  \\
&  &Date &Frequency& Resolution & \multicolumn{2}{c}{Beam} & Conversion \\
	         &	     &			        &     (GHz)    &       (km~s$^{-1}$)&  ($\arcsec$ $\times$ $\arcsec$) & PA ($\degr$)  & (K/(Jy~beam$^{-1}$)) \\
(1) & (2) & (3) & (4) &  (5) &  (6) &  (7) &  (8) \\
\tableline
1	& eSMA & 2009 March 22 & 337.671 &0.72 & 0.46 $\times$ 0.29 &52 & 80\\
2	& SMA\tablenotemark{b} & 2007 March 22      & 337.397 & 0.72 & 2.8 $\times$ 1.0 &52 & 3.7\\
3	& SMA $\&$ eSMA & -- & -- & 0.72 & 0.59 $\times$ 0.38 & 48 & 48\\
\tableline
\end{tabular}
\tablecomments{(1) Reference number of the corresponding data set. (2) Interferometer. (3) Observed date. (4) Rest Frequencies. (5) Spectral resolution. (6) $\&$ (7) Resulting synthesized beams. (8) Conversion factor of Jy~beam$^{-1}$ to K for each data set.}
\tablenotetext{a}{Set 1 was centered on coordinates ($\alpha_{J2000}$ = 16$^{h}$32$^{m}$22$\fs$898, $\delta_{J2000}$ = -24$\degr$28$\arcmin$35$\farcs$50). Sets 2 and 3 were centered on coordinates ($\alpha_{J2000}$ = 16$^{h}$32$^{m}$22$\fs$719, $\delta_{J2000}$ = -24$\degr$28$\arcmin$34$\farcs$30).} 
\tablenotetext{b}{\citet{Jorgensen:2011}.}
\end{center}
\end{table*}
One of the key question in studies of star and planet formation is when and how disk formation occurs. The formation of  a circumstellar disk, which will potentially result in planet formation, takes place during the rotating collapse of a dense pre-stellar core. Indeed, pure rotation accompanying collapse will give rise to a centrifugal disk, initially of low mass, evolving and growing with time \citep{Terebey:1984}. At the same time, the presence of a magnetic field can lead to the formation of a pseudo-disk around a young stellar object. The circumstellar disk is a product of relatively simple dynamics whereas the magnetic pseudo-disk arises through a magnetic pinch around  a young stellar object \citep{Basu:1998,Hennebelle:2009,Dapp:2010,Davidson:2011,Galli:1993a,Galli:1993}. 
A magnetic pseudo-disk grows continually as material is accreted and it can be much more massive and larger in the early stage of formation and evolution than the pure rotation disk \citep[e.g.][]{Basu:1998}. 
Such types of magnetic pseudo-disks have already been observed towards Class 0 young stellar objects  \citep[e.g. L1527 and IC3480-SMM2,][]{Davidson:2011,Ohashi:1997} and Class I sources  \citep[e.g. L1551 IRS 5 and HL~Tau,][]{Momose:1998,Takakuwa:2004,Lim:2006,Hayashi:1993}.
Observational studies of the kinematics of low-mass protostars can quantify the importance of rotation and magnetically modified infall, giving considerable insight  into the structure of Class 0 protostars and early disk formation in protostellar objects.

The well--studied deeply embedded low-mass protostar IRAS~16293-2422, which lies at a distance of 120~pc \citep{de-Geus:1989,Knude:1998,Loinard:2008} in the nearby  L1689N cloud located in the $\rho$ Ophiuchus cloud complex, is a potential source to undertake a kinematic study.
Two related components A and B \citep{Wootten:1989,Walker:1993}, hereafter IRAS16293A and IRAS16293B, separated by 5$\arcsec$ \citep[600~AU ;][]{Mundy:1992} are associated with this system. Although the nature of IRAS16293A as a protostellar object (Class 0) is commonly agreed upon, that of IRAS16293B is still debated: it could be a T Tauri star or an even younger protostellar object \citep[Class 0/I or candidate first hydrostatic core, e.g.][]{Stark:2004,Chandler:2005,Takakuwa:2007,Rao:2009,Pineda:2012,Loinard:2013,Zapata:2013}. The understanding of this region has been improved by high spatial resolution interferometric observations  of complex molecules including organic and prebiotic species for astrochemical studies and of simple species for dynamic and kinematic studies \citep{Kuan:2004,Huang:2005,Chandler:2005,Bottinelli:2004,Takakuwa:2007,Bisschop:2008,Jorgensen:2011,Jorgensen:2012,Pineda:2012}. In the present paper, we focus on the latter aspect.

The structure of the protostar IRAS~16293-2422 is complicated by the presence of infalling gas inside the circumstellar  envelop \citep{Walker:1986,Narayanan:1998,Ceccarelli:2000a,Chandler:2005,Takakuwa:2007}, as well as two outflows: one driven by IRAS16293A which is oriented in an east--west direction \citep[e.g. CO and SO observations, see][]{Mundy:1992,Yeh:2008,Jorgensen:2011} and a second that is oriented in a northeast-southwest direction \citep{Walker:1988,Mizuno:1990,Hirano:2001,Castets:2001,Garay:2002,Stark:2004}. Likewise, rotating material has also been observed towards this protostar \citep[e.g $^{13}$CO, SiO and C$^{18}$O observations see,][]{Mundy:1986a,Menten:1987,Mundy:1990,Zhou:1995,Schoier:2004,Huang:2005,Remijan:2006}. These studies have shown that the high angular resolution obtained with interferometers is required for detailed studies of the kinematics of low-mass protostars.

In this paper we investigate the kinematics of the molecular gas toward IRAS16293A with high angular resolution interferometric observations of carbon monoxide and monosulfide isotopologues (C$^{17}$O and C$^{34}$S, respectively). In Sect.~\ref{sec:Observations}, we present our Submillimeter Array \citep[SMA,][]{Ho:2004} and extended SMA \citep[eSMA,][]{Bottinelli:2008} data. A description of the data reduction and methodology used for combining these results is also given in this section. 
The basics results and the data analysis are presented in the sections \ref{sec:Results} and \ref{sec:Analysis}, respectively. 
The different scenarios which can explain the observed characteristics of the observations are discussed in Sect. \ref{sec:Analysis} with conclusions in Section \ref{sec:Conclusions}.

%
\section{Observations}
\label{sec:Observations}

\subsection{Extended SMA (eSMA) observations of carbon monoxide and monosulfide}

Observations of IRAS~16293-2422 were carried out with the eSMA on 2009 March 22 for 3.4~hours on source in single linear polarization mode.
The eSMA combines the SMA array (8 antennas of 6m), the James Clerk Maxwell Telescope (JCMT\footnote{The James Clerk Maxwell Telescope is operated by the Joint Astronomy Centre on behalf of the Science and Technology Facilities Council of the United Kingdom, the National Research Council of Canada, and (until 31 March 2013) the Netherlands Organisation for Scientific Research.}, 15m) and the Caltech Submillimeter Observatory (CSO\footnote{The Caltech Submillimeter Observatory is operated by Caltech under cooperative agreement with the National Science Foundation (AST-0838261).}, 10.4m) single-dish telescopes, yielding enhanced sensitivity and higher spatial resolution than the SMA alone.
The eSMA data  presented in this study cover one spectral setup at 337~GHz (see set 1 in Table 1). The phase-tracking centre was $\alpha_{J2000}$ = 16$^{h}$32$^{m}$22$\fs$898, $\delta_{J2000}$ = -24$\degr$28$\arcmin$35$\farcs$50. The correlator was configured for a single sideband, with a uniform spectral resolution over $\sim$2~GHz bandwidth divided into 24 \textquotedblleft chunks\textquotedblright , each of 104~MHz width and resulting in 128 channels. The weather conditions were good and stable and we estimate that the atmospheric opacity was 0.05-0.06.
Table \ref{tab1} presents the main parameters of the data (set 1).

In this paper, we focus only on emission lines of the carbon monoxide isotopologue C$^{17}$O (3-2) and the carbon monosulfide isotopologue C$^{34}$S (7-6). Table \ref{tab2} lists the spectroscopic parameters of these transitions.
To recover more extended emission we have combined the eSMA data (minimum baseline length of 32k$\lambda$)  with observations of the same lines from the SMA in its compact configuration  \citep[minimum baseline length of 11k$\lambda$, see][and set 2 in Table \ref{tab1}]{Jorgensen:2011}. Our combined eSMA and SMA observations are therefore not sensitive to structure extended on scales larger than 17$\arcsec$ \citep[see][]{Wilner:1994}.

%
\begin{table*}
\begin{center}
\caption{Spectroscopic line parameters of the carbon monoxide and carbon monosulfide isotopologues.\label{tab2}}
\begin{tabular}{l l l l l}
\tableline\tableline
Molecule\tablenotemark{a}& Frequency&Transition &$\langle$ S$\rm_{i,j}$$\mu$$^{2}$ $\rangle$& E$\rm_{u}$ \\
 & (MHz) && (D$^{2}$)& (K)   \\
 \tableline
C$^{17}$O & 337061.123& 3~-~2 & 0.01 & 32.35 \\
C$^{34}$S & 337396.459& 7~-~6 & 25.57 & 50.23 \\
\tableline
\end{tabular}
\tablenotetext{a}{All spectroscopic data from CO and CS isotopologues available from the CDMS molecular line catalog \citep{Muller:2001,Muller:2005} through the Splatalogue \citep[www.splatalogue.net,][]{Remijan:2007}  portal and are based on  laboratory measurements and model predictions by \citet{Klapper:2003,Cazzoli:2002,Goorvitch:1994,Winkel:1984,Ram:1995,Burkholder:1987,Gottlieb:2003,Ahrens:1999,Kim:2003,Bogey:1982,Bogey:1981}.}
\end{center}
\end{table*}

\subsection{Data Reduction}
\begin{figure}[h!]
\epsscale{1.0}
\plotone{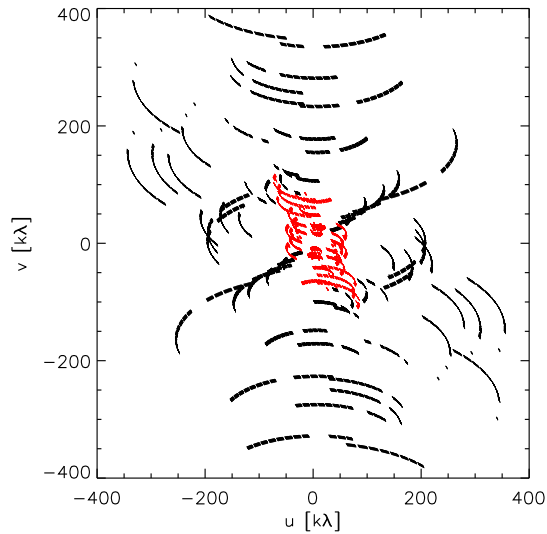}
\caption{Resulting $(u,v)$ coverage at 337~GHz  of the combined dataset (set 3 in Table \ref{tab1}) from the observed SMA tracks (black lines) with the observed eSMA tracks (red lines). \label{fg1}}
\end{figure}

The eSMA data were calibrated and reduced using the MIR/IDL package\footnote{https://www.cfa.harvard.edu/$\sim$cqi/mircook.html} \citep{Qi:2007}.
The nearby quasars 1626-298 and nrao530 (measured flux density of 1.3~Jy and 1.5~Jy, respectively) were used as phase and amplitude calibrators. The absolute flux calibration and the band-pass calibration were performed through observations of  the quasar 3C273, with an assumed flux of (10.3$\pm$1.5)~Jy, found by interpolating values obtained from the SMA calibrator list\footnote{http://sma1.sma.hawaii.edu/callist/callist.html} during the period February to April 2009. For details about the reduction of the SMA data see \citet{Jorgensen:2011}. The $(u,v)$ coverage of these datasets are shown in Fig.~\ref{fg1}.

%

Continuum subtraction and data imaging were performed using the MIRIAD software package \citep{Sault:1995}. The calibrated SMA and eSMA $(u,v)$ data were combined using the MIRIAD tasks \textit{uvaver} and \textit{invert} for both continuum and line analysis. Furthermore, in order to prevent any source position problems which could result from the different phase centres of the SMA and eSMA observations (see Table \ref{tab1}), the option \textit{mosaic} has been used. 
No cross flux calibration was done between the SMA and eSMA data. By comparing the continuum emission at different and overlapping baselines between the two datasets it appears that they are in agreement at the level of the calibration accuracy (20--30$\%$). Also, Our SMA continuum measurements (only) are in agreement with previous SMA observations at 305~GHz and 354~GHz by \citet{Chandler:2005} and  \citet{Kuan:2004} taking into account the different wavelengths and spatial frequencies covered by the observations. 

\subsection{Resulting continuum and molecular emission maps}
\begin{figure}[h!]
\epsscale{1.0}
\plotone{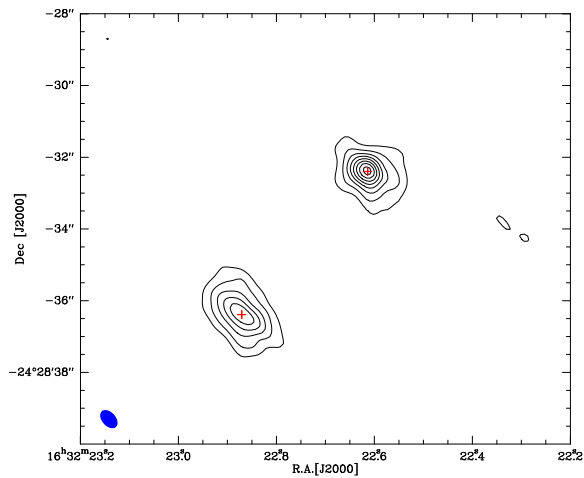}
\caption{Continuum maps obtained at 338.9~GHz toward IRAS~16293-2422 with the combined (LSB only) SMA and eSMA data. The synthesized beam is shown in the bottom left corner. The first contour and the level step are at 0.2~Jy~beam$^{-1}$. Red crosses indicate positions of the continuum sources IRAS16293A and IRAS16293B.\label{fg2}}
\end{figure}

Figure~\ref{fg2} shows the continuum emission observed at 338.9~GHz toward IRAS~16293-2422 with the combined (LSB only) SMA and eSMA data sets. 
The uniform weighted synthesized beam was 0.58$\arcsec$ $\times$ 0.38$\arcsec$ (P.A. of 47.6$\degr$).
From Gaussian fits in the $(u,v)$ plane the positions of the two main continuum sources, IRAS16293A and IRAS16293B, are: $\alpha_{J2000}$ = 16$^{h}$32$^{m}$22$\fs$87, $\delta_{J2000}$ = -24$\degr$28$\arcmin$36$\farcs$4 and $\alpha_{J2000}$ = 16$^{h}$32$^{m}$22$\fs$61, $\delta_{J2000}$ = -24$\degr$28$\arcmin$32$\farcs$4.
The structure of the emission of the continuum sources reveals both extended and compact emission. More specifically, continuum emission from IRAS16293A is clearly extended along a northeast-southwest axis, whereas toward IRAS16293B the emission appears more compact. 

%

The final combined C$^{17}$O and C$^{34}$S emission maps were restored using a uniform weighting, resulting in a synthesized beam size of 0.59$\arcsec$ $\times$ 0.38$\arcsec$ (P.A. of 47.3$\degr$ and 48.2$\degr$, respectively) which corresponds to $\sim$ 71 $\times$ 46~AU at a distance of 120~pc. The most important parameters of the combined data are listed in Table \ref{tab1} (see set 3).

%
\section{Results} 
\label{sec:Results}

In the following (sub)sections we will only present and discuss results on line emission that were obtained through  the combined SMA and eSMA data (see set 3 in Table \ref{tab1}).

%
\begin{figure*}
\epsscale{2.}
\plotone{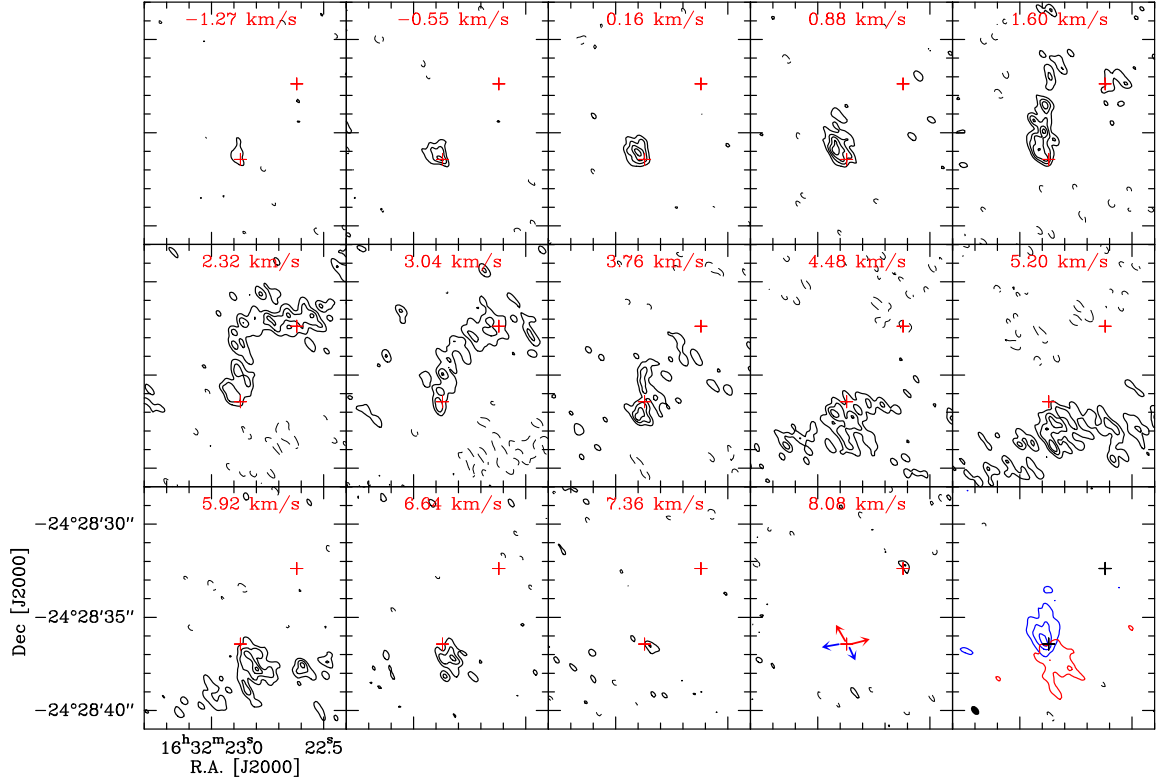}
\caption{Velocity channel maps of C$^{17}$O towards IRAS~16293-2422. The $v$$\rm_{LSR}$ velocity is indicated on each plot. The first contour is at 3$\sigma$ and the level step is 2$\sigma$ (1$\sigma$ level is 123~mJy~beam$^{-1}$). Red crosses indicate positions of the continuum sources IRAS16293A and IRAS16293B (see Sect. \ref{sec:Results}.1). Principle red and blue-shifted directions of the two outflows arising from source A are indicated in the 8.08~km/s channel map \citep{Mundy:1992,Yeh:2008,Jorgensen:2011,Walker:1988,Mizuno:1990,Hirano:2001,Castets:2001,Garay:2002,Stark:2004}. The bottom right panel shows the integrated blue- and red-shifted emission map of C$^{17}$O. The blue-shifted emission is integrated over the velocity channels from $v$$\rm_{LSR}$= $-$3.0 to 3.8~km~s$^{-1}$ and the red-shifted emission between 3.8 and 9.0~km~s$^{-1}$. The first contour and the level step are 1.4~Jy~beam$^{-1}$~km~s$^{-1}$. The SMA $\&$ eSMA synthesized beam is 0.59$\arcsec$ $\times$ 0.38$\arcsec$ (see Table~\ref{tab1}).\label{fg3}}
\end{figure*}

%
\begin{figure*}
\epsscale{2.0}
\plotone{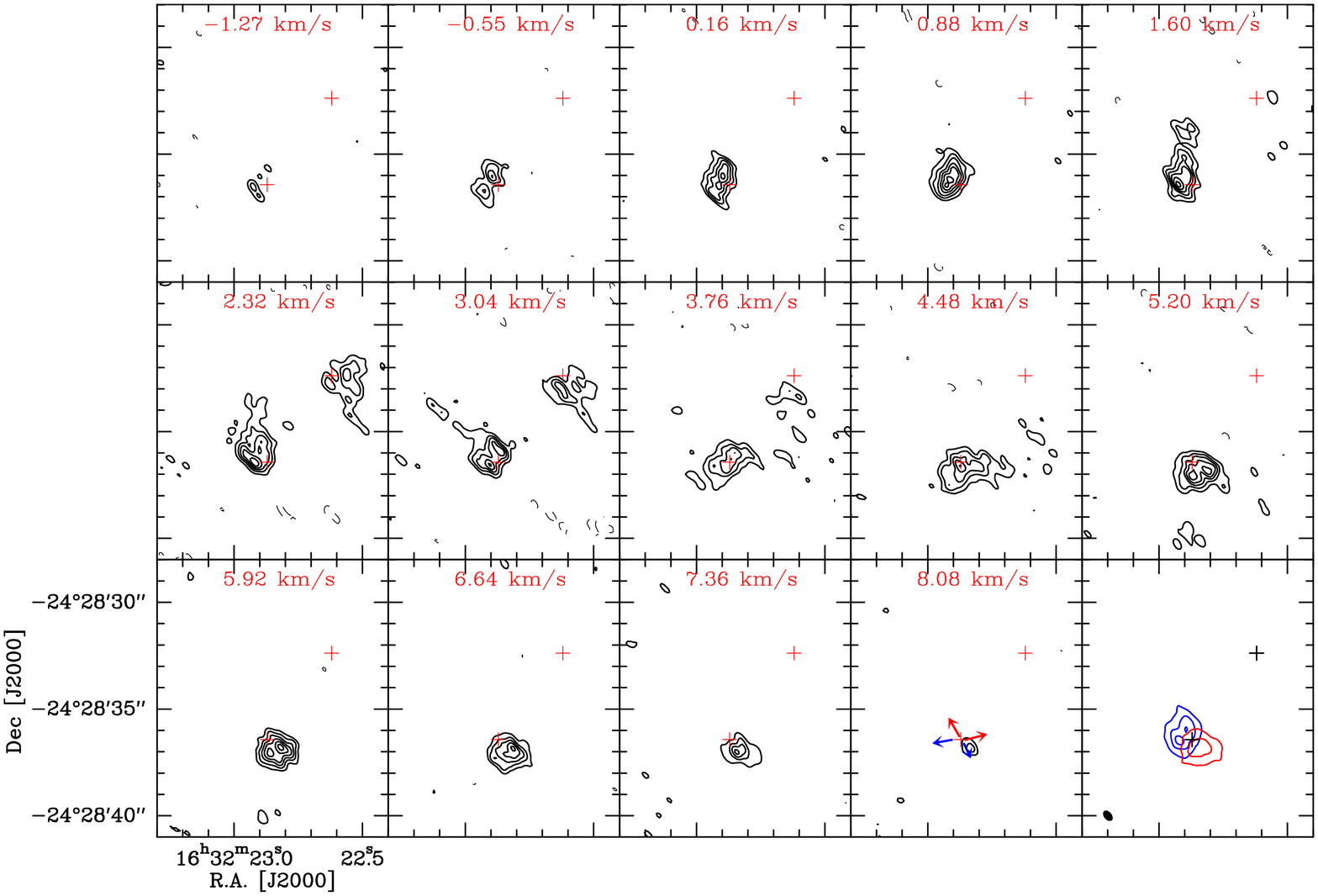}
\caption{Velocity channel maps  of C$^{34}$S towards IRAS~16293-2422. The $v$$\rm_{LSR}$ velocity is indicated on each plot. The first contour  is at 3$\sigma$ and the level step is 2$\sigma$ (1$\sigma$ level is 116~mJy~beam$^{-1}$). Principle red and blue-shifted directions of the two outflows arising from source A are indicated in the 8.08~km/s channel map \citep{Mundy:1992,Yeh:2008,Jorgensen:2011,Walker:1988,Mizuno:1990,Hirano:2001,Castets:2001,Garay:2002,Stark:2004}. The bottom right panel shows the integrated blue- and red-shifted emission map of C$^{34}$S. The blue-shifted emission is integrated over the velocity channels from $v$$\rm_{LSR}$= $-$3.0 to 3.8~km~s$^{-1}$ and the red-shifted emission between 3.8 and 9.0~km~s$^{-1}$. The first contour and the level step are 1.9~Jy~beam$^{-1}$~km~s$^{-1}$. The SMA $\&$ eSMA synthesized beam is 0.59$\arcsec$ $\times$ 0.38$\arcsec$.\label{fg4}}
\end{figure*}

\subsection{Emission maps and velocity structure }

\begin{figure}
\includegraphics[angle=270,width=7.5cm]{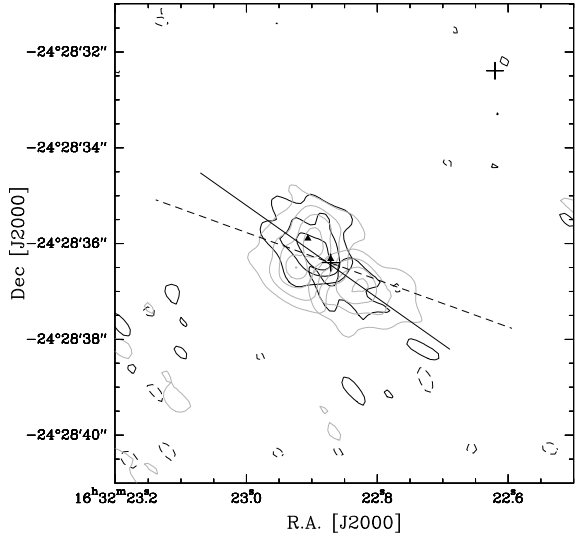}
\caption{Integrated C$^{17}$O (black) and C$^{34}$S (grey) emission maps at the higher blue-shifted and red-shifted velocities. The blue-shifted emission is integrated over the velocity channels from $v$$\rm_{LSR}$= $-$2.6 to 0.7~km~s$^{-1}$ and the red-shifted emission between 6.6 and 9.0~km~s$^{-1}$. The first contour and the level step are at 2$\sigma$ (with 2$\sigma$ = 0.7 and 0.6~Jy~beam$^{-1}$~km~s$^{-1}$ for the blue- and red-shifted emission for C$^{17}$O and 0.6 and 0.5~Jy~beam$^{-1}$~km~s$^{-1}$ for the blue- and red-shifted emission for the C$^{34}$S). The full black line indicates the orientation of the northeast-southwest gradient (P.A. $\sim$54$\degr$) whereas the dash line indicates the NE-SW outflow orientation on small scale  (P.A. $\sim$70$\degr$) as reported by \citet{Yeh:2008}. Crosses and filled-triangles indicates positions of the sources IRAS16293A and IRAS16293B and,  sources Aa and Ab \citep{Chandler:2005}, respectively.\label{fg5}}
\end{figure}

Figures \ref{fg3} and \ref{fg4} show \textit{i)} the channel maps of the C$^{17}$O (3-2) and C$^{34}$S (7-6) from $v$$\rm_{LSR}$ = $-$1.3~km~s$^{-1}$ to 8.1~km~s$^{-1}$, respectively and,  \textit{ii)} the integrated emission maps of these species. 
The detailed structure of the C$^{17}$O (3-2) and C$^{34}$S (7-6) line emission is complex, showing a velocity gradient oriented in a northeast-southwest direction with respect to IRAS16293A. Indeed, the C$^{17}$O and C$^{34}$S velocity channel maps, presented in Figs. \ref{fg3} and \ref{fg4}, show that:
\begin{itemize}
\item from $v$$\rm_{LSR}$= $-$0.6~km~s$^{-1}$ to 3.8~km~s$^{-1}$ the blue-shifted emission around the systemic velocity peaks toward the north/northeast of IRAS16293A,
\item at  $v$$\rm_{LSR}$ of  2.3~km~s$^{-1}$ and 3.0~km~s$^{-1}$, some emission appears around  IRAS16293B,
\item from $v$$\rm_{LSR}$= 1.6~km~s$^{-1}$ to 5.9~km~s$^{-1}$, the C$^{17}$O channel maps present some elongated features along an east-west direction, which are consistent with the distribution of the SiO (8-7) emission observed toward IRAS~16293-2422 by \citet{Jorgensen:2011},
\item and from $v$$\rm_{LSR}$= 3.8~km~s$^{-1}$ to 7.4$-$8.1~km~s$^{-1}$ the red-shifted emission clearly peaks toward the south/southwest of IRAS16293A.
\end{itemize}
Although the channel maps are complex, the bulk of the C$^{17}$O  and C$^{34}$S emission is associated with the red and blue structures seen in the integrated intensity emission maps (see final panels of figs. \ref{fg3} and \ref{fg4}).
Figure~\ref{fg5} shows the higher red- and blue-shifted integrated emission maps of both isotopologues from $v$$\rm_{LSR}$=6.6 to 9.0~km~s$^{-1}$ and  $v$$\rm_{LSR}$=$-$2.6 to 0.7~km~s$^{-1}$, respectively. The orientation northeast- southwest (NE-SW) of the velocity gradient is clearly seen in Fig.~\ref{fg5}. The resulting measured position angle is $\sim$54$\degr$.

%

%
\subsection{Spectra}

\begin{figure*}
\epsscale{4.}
\plottwo{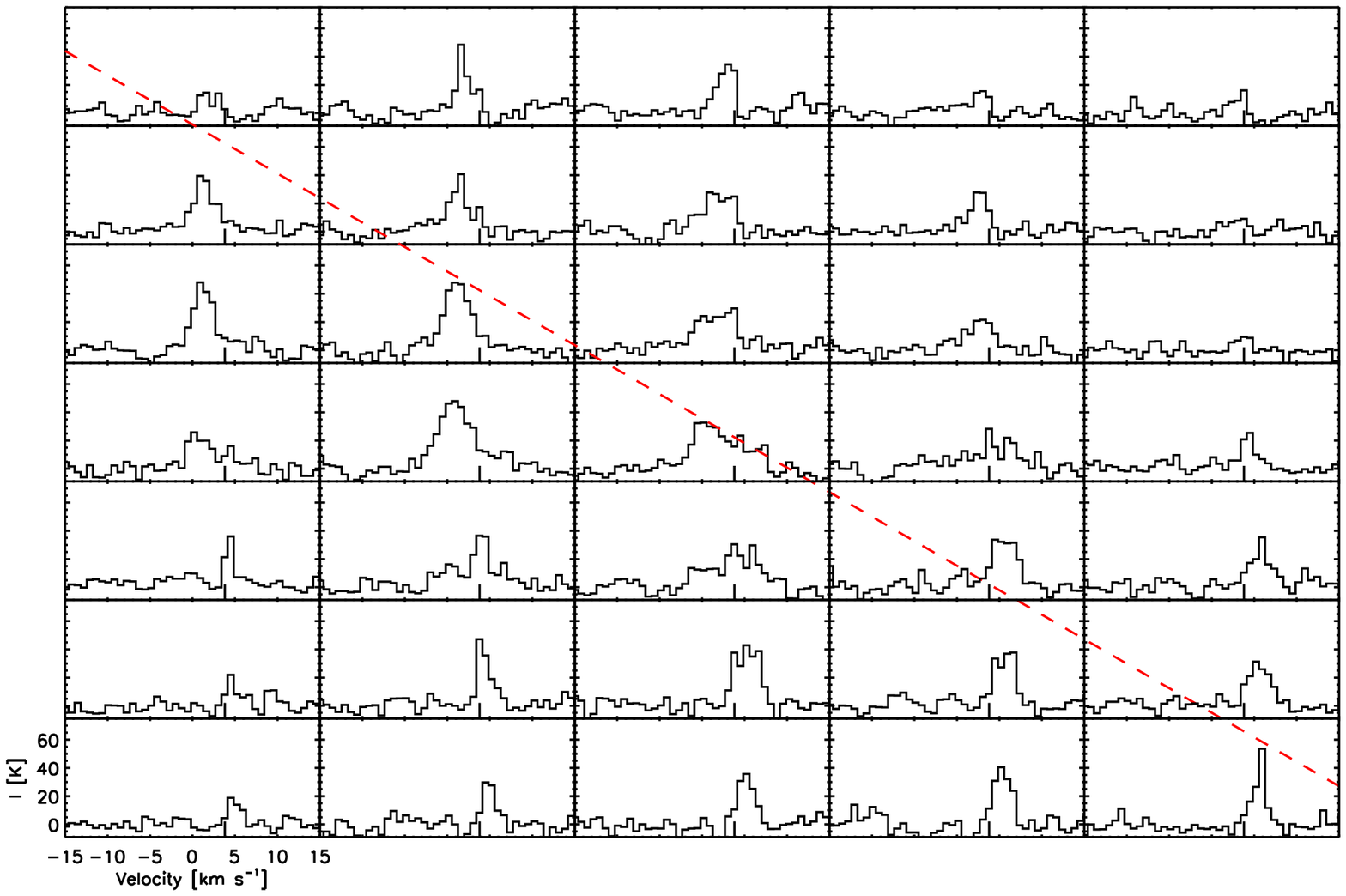}{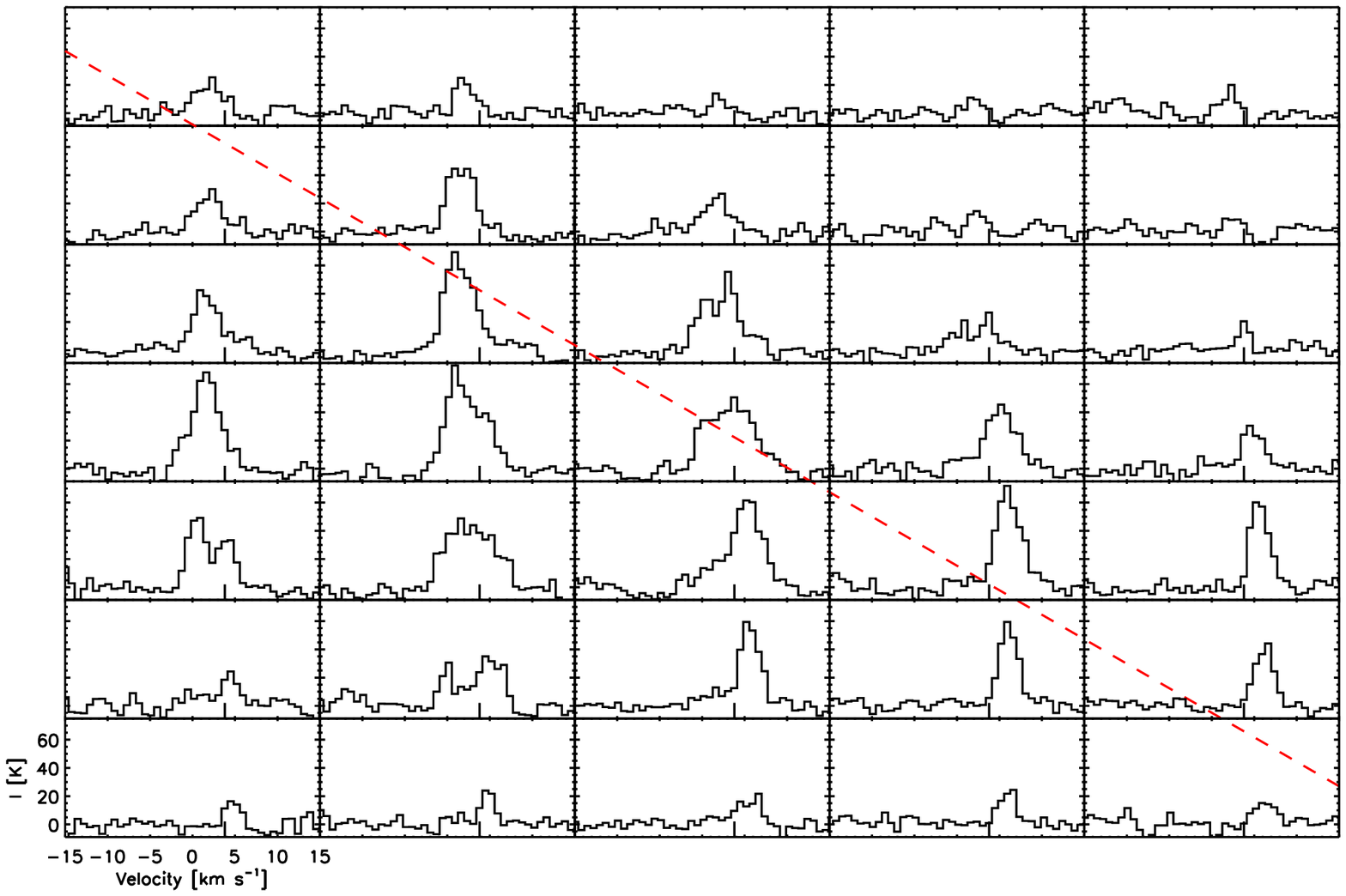}
\caption{C$^{17}$O (top) and C$^{34}$S (bottom) spectral maps displayed on a (R.A, Dec) grid centering on IRAS16293A. The spectra are at intervals of 0.5$\arcsec$ and are integrated over an area of 0.5$\arcsec$. The red dashed line indicates the northeast-southwest direction of the observed velocity gradient. In each spectrum panel, the short black  solid line indicates $v$$\rm_{LSR}$=3.8~km~s$^{-1}$. \label{fg6}}
\end{figure*}

Fig.~\ref{fg6} displays the spectral profiles of the  C$^{17}$O  and C$^{34}$S, on a (R.A. , Dec) grid centered on IRAS16293A. 
Most of the blue-shifted emission is stronger in the northern/northeast offsets of IRAS16293A whereas the red-shifted emission is stronger in the southern/southwest offsets. 

%

Toward the central source (IRAS16293A), both C$^{34}$S  and C$^{17}$O spectra can approximately be described by a single Gaussian ($\Delta$v$_{1/2}$ of $\sim$7--7.7~km~s$^{-1}$) centered at $\sim$3.4~km~s$^{-1}$ for C$^{34}$S, which is close to the systemic velocity of the cloud  \citep[3--4~km~s$^{-1}$, see][]{Mizuno:1990,Jorgensen:2011} and, at $\sim$1.4~km~s$^{-1}$ for C$^{17}$O. The C$^{17}$O spectrum being spread toward source IRAS16293A (Fig.~\ref{fg6}), the resulting fit is poor, that leads to a non-accurate determination of the center of the Gaussian.

%
\section{Analysis and discussion} 
\label{sec:Analysis}

\subsection{Missing flux}

The present section aims to estimate the portion of the total flux resolved out by the interferometer.
To estimate the fraction of the total flux is missing, we compared the SMA and eSMA data to archival JCMT observations\footnote{The JCMT data used here are public and available from the JCMT Science Archive portal, see http://www.jach.hawaii.edu/JCMT/archive/.} and to published CSO observations \citep{Blake:1994} .

The SMA and eSMA C$^{17}$O spectrum was convolved with a Gaussian beam to mimic the JCMT beam at 337~GHz (15$\arcsec$), and  the JCMT spectrum has been converted into main beam temperature (T$\rm_{mb}$) using T$\rm_{mb}$=T$\rm_{A}^{*}$/$\eta\rm_{mb}$, where T$_{A}^{*}$ is the antenna temperature and $\eta\rm_{mb}$ the main beam efficiency. We have adopted a value of 0.64 for $\eta\rm_{mb}$ \citep{Buckle:2009}. In addition, the JCMT spectrum has been smoothed to the same spectral resolution (0.72~km~s$^{-1}$) as that of the combined SMA and eSMA spectrum (see Table 1).

\begin{figure}[h!]
\epsscale{1.0}
\plotone{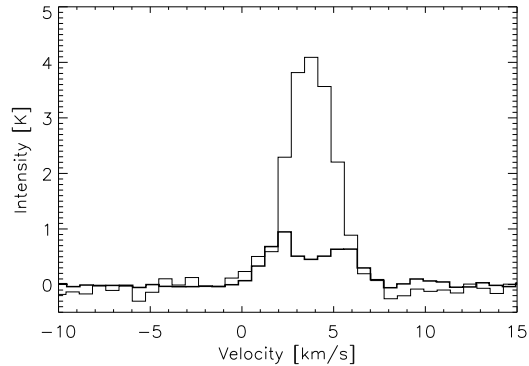}
\caption{C$^{17}$O emission spectra toward IRAS~16293-2422. The grey histogram shows the JCMT spectrum and the black histogram illustrates the combined SMA/eSMA spectrum convolved with the JCMT beam at 337~GHz (15$\arcsec$).\label{fg7}}
\end{figure}

At the systemic velocity of the cloud  \citep[3 - 4~km~s$^{-1}$,][]{Mizuno:1990,Jorgensen:2011} almost all the C$^{17}$O emission is resolved out, since this emission is present largely in the extended surrounding cold gas. However, in the line wings (from $v$$\rm_{LSR}$ = 0~km~s$^{-1}$ to 2~km~s$^{-1}$ and from $v$$\rm_{LSR}$ = 5.5~km~s$^{-1}$ to 7.5~km~s$^{-1}$, as shown in Fig. \ref{fg7}), 60 \%--70 \% of the C$^{17}$O flux  is recovered by the combined SMA and eSMA observations.

Concerning the C$^{34}$S emission, no single-dish spectra are available. We therefore compared the integrated line flux, reported by \citet{Blake:1994} from CSO observations, with the integrated line flux derived from the convolution of the SMA and eSMA C$^{34}$S spectrum with a Gaussian beam similar to the CSO beam at 337~GHz (20$\arcsec$). The comparison shows that 59$\pm$5 \% of  the C$^{34}$S emission is resolved out. We conclude that C$^{34}$S emission is filtered out by the interferometer in a manner to an extent similar to that  for C$^{17}$O emission.

%

\subsection{Interpretation of the velocity data for C$^{17}$O and C$^{34}$S}
\label{sectionpv}
%

%
\begin{figure*}
\includegraphics[angle=270,width=8.cm]{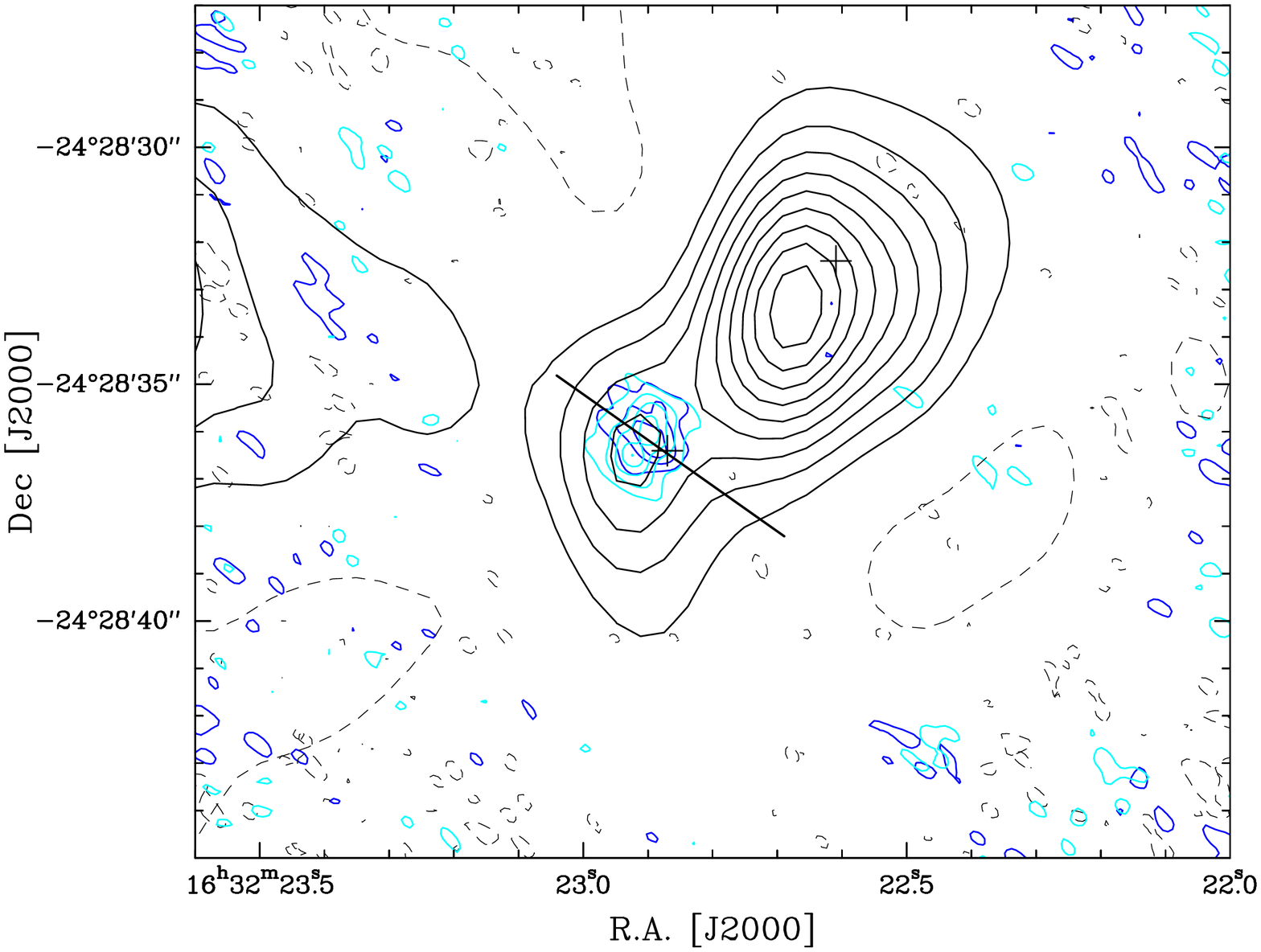}
\includegraphics[angle=270,width=8.cm]{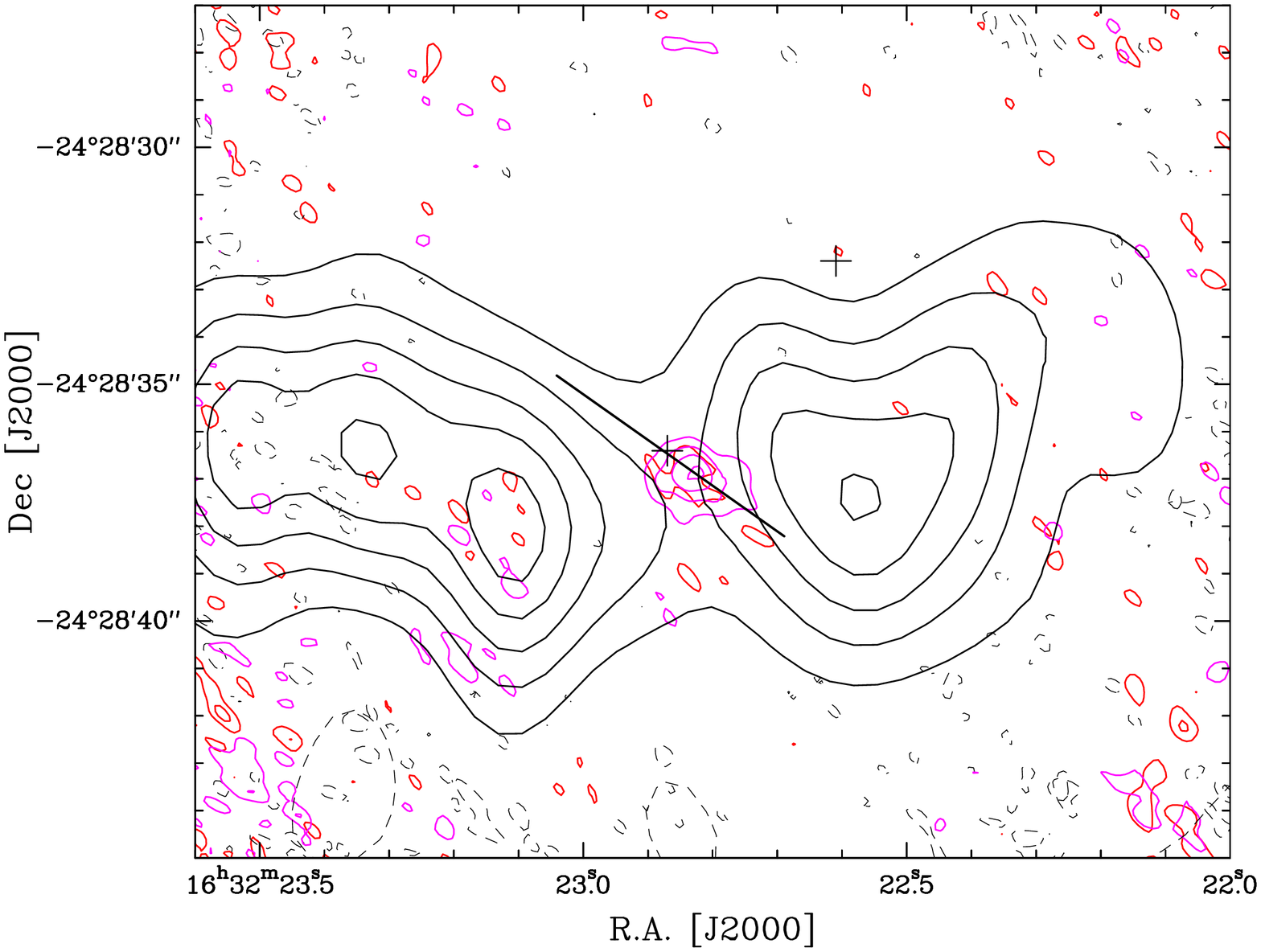}
\caption{
{\em Left:} Integrated C$^{17}$O (blue), C$^{34}$S (cyan) and CO (dark) emission maps at the higher blue-shifted velocities (from $v$$_{LSR}$ in the range $-$2.6 to 0.7~km~s$^{-1}$). The first contour and the level step are at 2$\sigma$ (with 2$\sigma$=0.7, 0.6 and 5~Jy~beam$^{-1}$~km~s$^{-1}$ for C$^{17}$O, C$^{34}$S and CO, respectively.
{\em Right:} Integrated C$^{17}$O (red), C$^{34}$S (magenta) and CO (dark) emission maps at the higher red-shifted velocities (from $v$$_{LSR}$ in the range 6.6 to 9.0~km~s$^{-1}$). The first contour and the level step are at 2$\sigma$ (with 2$\sigma$=0.6, 0.5 and 9~Jy~beam$^{-1}$~km~s$^{-1}$ for C$^{17}$O, C$^{34}$S and CO, respectively.
The CO (2--1) observations were carried out with the SMA by \citet{Jorgensen:2011}. Crosses indicates positions of the sources IRAS16293A and IRAS16293B. Black line indicates the orientation of the C$^{17}$O and C$^{34}$S northeast-southwest gradient (P.A. $\sim$54$\degr$, see Fig.~\ref{fg5}).\label{fg8}} 
\end{figure*}

\subsubsection{Indication of non-outflowing gas}

The purpose of this section is to investigate the hypothesis that C$^{17}$O and C$^{34}$S show any velocity gradient in the propagation direction of the outflow.

Although the orientation of northeast-southwest  velocity gradient seen in the C$^{17}$O and C$^{34}$S channel and spectral maps (Figs. \ref{fg3}, \ref{fg4} and \ref{fg6}) is aligned with the one of the northeast-southwest (NE--SW) outflow of the source \citep[P.A. of $\sim$45$\degr$,][]{Walker:1988,Mizuno:1990,Hirano:2001,Castets:2001,Garay:2002,Stark:2004,Chandler:2005,Loinard:2013}, the NE-SW outflow harbors large scale structures only ($\sim$15 000~AU) and does not show any small scale structures ($\sim$3 000~AU), as discussed in \citet{Loinard:2013} and \citet{Yeh:2008}, in contrast to C$^{17}$O and C$^{34}$S that are only probing small scales.

For the east-west (E--W) outflow, \citet{Yeh:2008} showed that it had a complex structure in CO emission -- but on small scales is oriented in the NE--SW direction with a position angle of about 70$\degr$. \citet{Takakuwa:2007} found that HCN emission observed by the SMA was partly due to the E--W outflow.
Figure~\ref{fg8} shows  the integrated high blue-shifted ($v$$_{LSR}$=$-$2.6 to 0.7~km~s$^{-1}$) and red-shifted ($v$$_{LSR}$=6.6 to 9.0~km~s$^{-1}$) velocities C$^{17}$O and C$^{34}$S and CO (2--1)\footnote{The CO (2--1) observations were carried out with the SMA by \citet{Jorgensen:2011}.}. Contrary to HCN emission, the C$^{17}$O and C$^{34}$S emission appears to be nearly $\sim$40$\degr$ and 90$\degr$ different in angles than the CO emission for the high red-shifted and blue-shifted emission, respectively. Furthermore, as shown in Fig.~\ref{fg5}, a position angle of 70$\degr$ doesn't fit on the higher red- and blue-shifted integrated emission maps of both isotopologues. Our finding suggests that C$^{17}$O and C$^{34}$S are unlikely to probe a structure which is associated with the east-west outflow and, could originate from a different source than IRAS16293A, which likely drives the E-W outflow \citep{Yeh:2008}. 

\subsubsection{Rotation pattern} 

The emission of C$^{17}$O and C$^{34}$S, which we assume to be optically thin, is  complex. The observed northeast/north -- southwest/south velocity gradients and line profiles do not appear to trace outflowing material but may indicate rotation signatures. The presence of rotating material towards source IRAS16293A \citep[roughly perpendicular to the second outflow of IRAS16293, that is oriented in an east--west direction, see][]{Mundy:1992,Yeh:2008,Jorgensen:2011} has been reported based on single-dish and interferometric observations of $^{13}$CO, C$^{18}$O, H$_{2}$CO and C$^{32}$S \citep{Mundy:1986a,Menten:1987,Mundy:1990,Zhou:1995,Schoier:2004}. Likewise, from SMA observations of HCN and HC$^{15}$N, \citet{Takakuwa:2007} and \citet{Huang:2005} also reported a velocity gradient in a northeast/north-southwest/south direction (i.e. along the outflow oriented NE-SW). \citet{Takakuwa:2007} interpreted the observed  flattened structure as an accreting disk and \citet{Huang:2005} suggested the emission is probing  an inclined (30$\degr$, with respect to the sky) rotating circumstellar disk.  These earlier velocity gradient observations are all consistent with our SMA and eSMA C$^{17}$O and C$^{34}$S observations (see Figs.~\ref{fg3}, \ref{fg4}, \ref{fg5} and  \ref{fg6}), but are of lower resolution.

Rotational motion, in particular of Keplerian type, can be distinguished from solid body motions and infall signatures through position-velocity diagrams (hereafter PV-diagrams).
Typically, if the gas is dominated by rotation, the PV-diagram along the supposed axis of rotation should present no evidence of rotation, whereas the PV-diagram along the perpendicular axis should show the maximum effect \citep[e.g.][]{Brinch:2009}.
Figure~\ref{fg9} presents the PV-diagrams for C$^{17}$O and C$^{34}$S centered on the position of IRAS16293A \textbf{ \textit{i)}} for a slice along the northeast-southwest velocity gradient direction ($\sim$54$\degr$) and \textbf{ \textit{ii)}} for a slice along its perpendicular direction ($\sim$144$\degr$), which is assumed to be the rotational axis. 
We note that, for both isotopologues, no evidence of systemic motions is observed along the supposed rotational axis. Also, the perpendicular axis, that is oriented in the northeast-southwest direction, clearly represents a strong rotation pattern (see the left hand upper and middle panels of Fig. \ref{fg9}): \textit{i)} the blue-shifted emission is located in the north whereas the red-shifted emission is mainly seen in the south, \textit{ii)} the related main blue-and red-shifted emission peaks are shifted west and east from the systemic velocity axis and, \textit{iii)} the emission drops at low velocities and the distribution of the emission can be described by a \textquotedblleft butterfly wing\textquotedblright \  shape in the upper-left and bottom-right quadrants only.
The positions of the blue and red-shifted emission peaks in the C$^{17}$O and C$^{34}$S velocity profiles are consistent with PV-diagrams in CS, $^{13}$CO, C$^{18}$O, HCN and HC$^{15}$N, towards IRAS16293A, for which rotation of material has been reported \citep[see][]{Mundy:1986a,Mundy:1990,Zhou:1995,Menten:1987,Huang:2005}.

\subsubsection{Keplerian-type rotation or reflection of a rotating/infalling core ?}

\begin{figure*}
\epsscale{	2.2}
\plotone{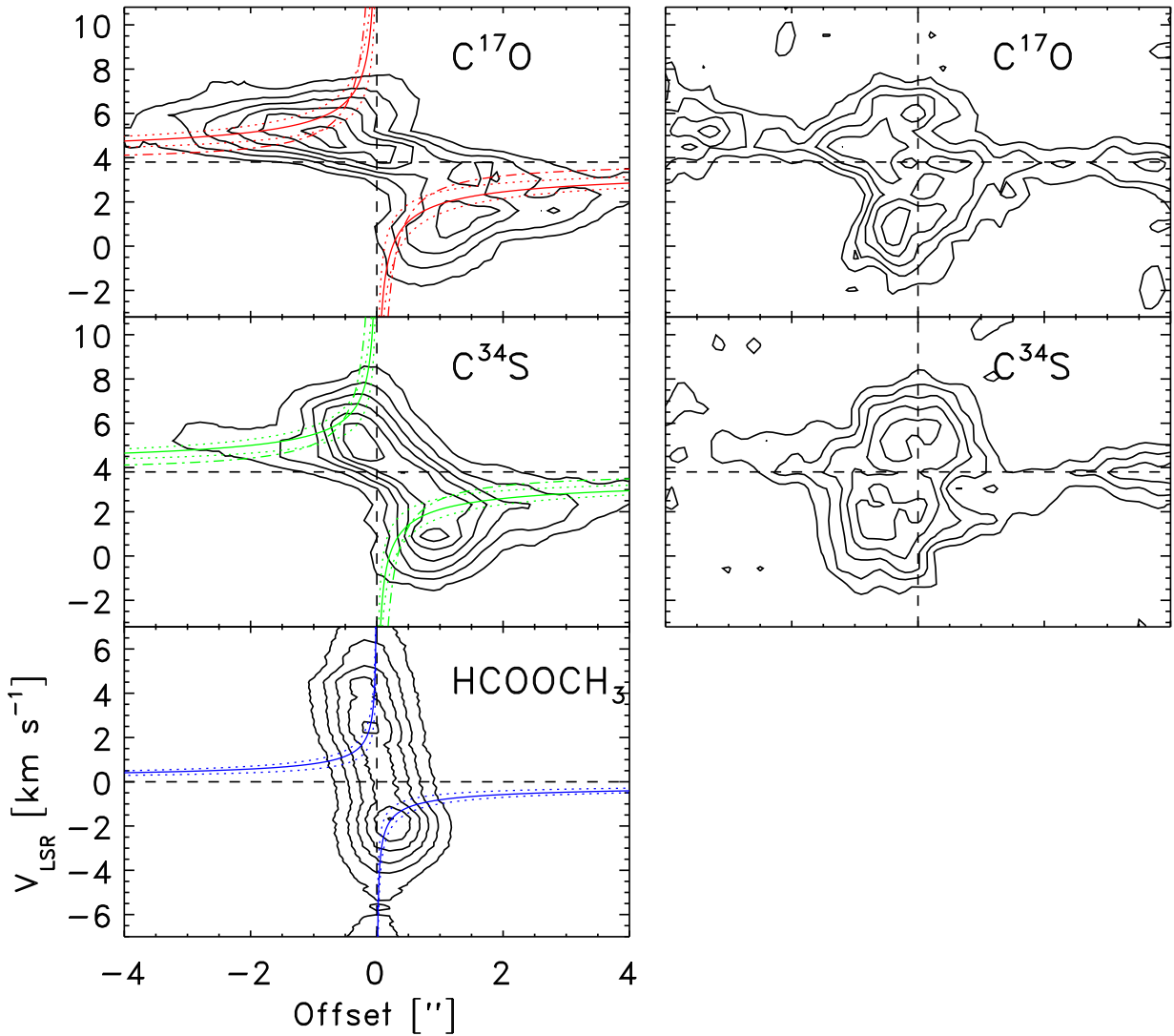}
\small{
\caption{Upper and middle panels: PV diagrams respectively in C$^{17}$O  and C$^{34}$S centering on the IRAS16293A position at $v$$\rm_{LSR}$=3.8~km~s$^{-1}$. Left-hand panels correspond to a slice perpendicular to the supposed axis of rotation ($\sim$54$\degr$) and right-hand panels correspond to a slice in the direction of the rotation axis ($\sim$144$\degr$). The first contour and level steps are at 3.36~Jy.beam$^{-1}$ for C$^{17}$O and at 3.96~Jy.beam$^{-1}$ for C$^{34}$S. Bottom left panel: PV diagram in HCOOCH$_{3}$ (transition at 220.166GHz) toward IRAS16293A corresponding to the direction of the NE-SW velocity gradient \citep[ALMA-SV data, see][]{Pineda:2012}. Over-plotted are the predicted curves for purely Keplerian rotation around a 0.49~M$_\odot$ central object (solid red lines, top panel), a 0.39~M$_\odot$ central object (solid green lines, middle panel) and a 0.09~M$_\odot$ central object (solid blue lines, bottom panel) ;  as well as predictions (dotted lines) for a  $\pm$50$\%$ uncertainty on the mass. In addition, 1$/$r rotation curves are over-plotted, in chained, in the C$^{17}$O and  C$^{34}$S PV diagrams (upper and middle left--hand panels). \label{fg9}}}
\end{figure*}

Both C$^{17}$O and C$^{34}$S  PV-diagrams present a \textquotedblleft butterfly wing\textquotedblright \  shape along the northeast-southwest axis. This specific pattern is usually associated with Keplerian motion of the gas. Indeed, it has been seen toward several Class I young stellar objects for which disks in Keplerian rotation have been observed \citep[e.g L1489 IRS, IRS43, IRS 63, Elias 29 and HH~111, see][] {Hogerheijde:2001,Lommen:2008,Jorgensen:2009,Lee:2010}. Our results in Fig. \ref{fg9} appear to indicate that motion of the gas could be dominated by Keplerian-type rotation. 
Nonetheless, rotation of material which has a constant angular momentum could also fit the observed patterns. 
In order to estimate whether the rotation is purely Keplerian or reflecting a rotating infalling core, simple models of a rotation velocity profile have been performed (see left--hand panels in Fig. \ref{fg9}). The velocity field was parameterized by a rotational velocity depending on the radius:
\begin{itemize}
\item for purely Keplerian rotation, where the stellar mass dominates over the envelope mass, we adopted a velocity profile for a disk seen edge-on ;
\begin{equation} \rm
V = \sqrt{\frac{GM_{*}}{r}},
\end{equation}
where $M\rm_{*}$ the mass of the central object,
\item and for infall with conservation of the angular momentum, we used a simple power law ; V $\sim$ r$^{-1}$ assuming an angular momentum of 150~AU~km~s$^{-1}$, that is in agreement with the typical values reported by \citet{Belloche:2013}.
\end{itemize}
In addition, our velocity profile studies also includes gas probed by the methyl formate molecule (HCOOCH$\rm_{3}$). \citet{Pineda:2012} have suggested, from ALMA science verification observations, that HCOOCH$\rm_{3}$ PV-diagram along the north/northeast--south/southwest direction toward IRAS16293A is consistent with rotation of a disk.

The best models, roughly fitting both emission peaks and the 3$\sigma$ edge, are presented in the left hand panels on Fig. \ref{fg9}. For the infall model, our best model is well described by the following law: $\rm $V$~=~1.5({\frac{r}{100~AU}})^{-1}$~km~s$^{-1}$. A salient result is that Keplerian rotation cannot be unambiguously distinguished from rotation conserving its angular momentum the C$^{17}$O and C$^{34}$S PV-diagrams. 
Our data are therefore consistent with rotation and we might be observing a  change of rotation profile in the envelope as observed in some Class I objects \citep[e.g.][]{Lee:2010,Momose:1998} but  we can make no firm conclusion.

Also, the rotation seems to reflect the decrease in envelope mass. Indeed, the predicted curves for a purely Keplerian rotation profile are obtained for a central object of 0.49~M$_{\sun}$ based on C$^{17}$O observations -- which is in agreement within 10$\%$ with the masses of the central object derived by \citet{Looney:2000,Huang:2005,Pineda:2012}, but a factor 2 lower than the central mass derived in \citet{Takakuwa:2007} -- and,  for central objects of 0.39~M$_{\sun}$ and  0.09~M$_{\sun}$ from C$^{34}$S and HCOOCH$\rm_{3}$ observations, respectively. The results indicate that a single central mass is inconsistent with the data. One possibility is that  the envelope mass (M$\rm_{env}$(r)) is getting closer to the stellar mass resulting in a rotation profile better described by: 
\begin{equation} \rm
 V \varpropto \sqrt{(G(M_{*}+M_{env}(r)))/ r}.
 \end{equation}
Also, the HCOOCH$_{3}$ ALMA-SV observations suggest that if the mass of the central object is greater than 0.1~M$_\odot$ then a purely Keplerian velocity field will be inconsistent with our measurements  (see Fig. \ref{fg9}).

\begin{figure*}
\epsscale{1.5}
\plotone{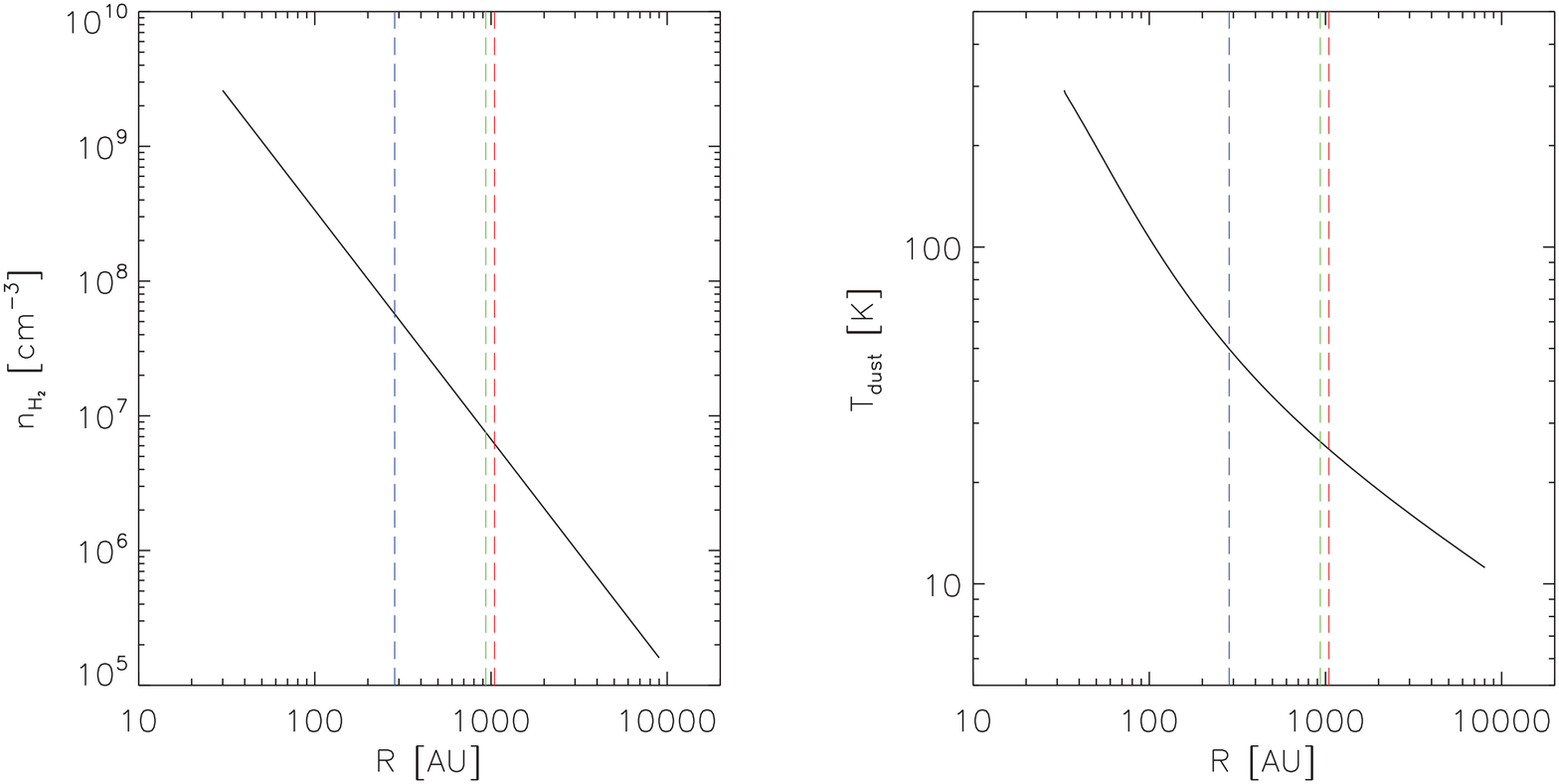}
\caption{Density (left panel) and dust temperature (right panel) profiles of IRAS 16293-2422 taken from \citet{Schoier:2002}. Over-plotted, in dashed lines, are the radii equivalent to the radii of the enclosed masses corresponding to the HCOOCH$_{3}$ (blue), C$^{34}$S (green) and C$^{17}$O  (red) mass estimates.\label{fg10}}
\end{figure*}

In summary, our analysis suggests that  the velocity field is inconsistent with pure Keplerian rotation around a single point-mass ; rather the enclosed envelope mass plus stellar mass is influencing the distribution of the rotational velocities. In this instance, different tracers may have different enclosed masses and thus the rotation curves may be more distinct. In this case, the inferred dynamical mass from C$^{17}$O and C$^{34}$S must be larger than the dynamical mass from HCOOCH$_{3}$ with each probing larger scales. This point is illustrated in Figure \ref{fg10}, which shows the density and dust temperature profiles of IRAS 16293-2422  from \citet{Schoier:2002} and indicates the radii equivalent to the radii of the enclosed masses corresponding to the HCOOCH$_{3}$, C$^{34}$S and C$^{17}$O mass estimates. We also conclude that methyl formate is clearly probing denser and warmer gas than C$^{34}$S and C$^{17}$O.

%

%

%
\subsubsection{Other possible hypotheses} 
\label{sec:discussion}

Our analysis shows that C$^{34}$S and C$^{17}$O are probing a rotating structure. Nevertheless, other scenarios, that cannot be ruled out at the present time, can also explain our observations.

The first hypothesis implies that the observed structure may be contaminated by infall motions in the envelope. In that connection, \citet{Takakuwa:2007} interpreted the observed flattened structure seen in HCN and HC$^{15}$N, which shows a velocity gradient along a northeast/southwest direction, but with a different P.A., as an accreting disk. Although, our present data are consistent with rotation, we cannot rule out the possibility that a part of the material is actually infalling at some place in the envelope (i.e. at P.A.  other than 144$\degr$). In that light, \citet{Tobin:2012} have shown that at scale larger than 1000~A.U. a mix of infall and (solid-body) rotation can result in a PV-diagram which present similarities with the PV-diagrams for C$^{17}$O and C$^{34}$S (see, for example, Figure 1 of \citealp{Tobin:2012}). Likely, infall can also affect a Keplerian-type PV-diagram on scale smaller than 1000 AU.

Another hypothesis involves the nature of the circumstellar infalling envelope. Recently, \citet{Tobin:2011,Tobin:2012} have shown that the morphology of an envelope can affect the kinematics on scales larger than 1000~AU.  Indeed, due to projection effects, a filamentary infalling envelope could give rise to a PV-diagram similar to a differential rotation PV-diagram. Although unlikely, the nature of the envelope might affect the kinematic we are observing at scales close to 1000~AU (see Fig.~\ref{fg10}).

Alternatively, C$^{17}$O and C$^{34}$S could be probing a magnetic pseudo-disk \citep[see][]{Galli:1993a,Galli:1993,Davidson:2011,Hennebelle:2009}. In this connection, \citet{Davidson:2011} have shown that pseudo-disks are observed for Class 0 young stellar objects (e.g. L1527, IC348--SMM2). 
The salient reasons which support the hypothesis that a pseudo-disk may give rise to our observations are as follows:
\begin{itemize}
\item observational data suggest the presence of a large flattened infalling and rotating structure in the inner part of the envelope at radii less than 8000~AU \citep[see Fig.~\ref{fg10} and][]{Schoier:2002},
\item polarization observations support the presence of a magnetic pseudo-disk.
\end{itemize}
With regard to the magnetic field, large scale polarization has been reported by \citet{Rao:2009} and \citet{Tamura:1995} based on observations of the dust continuum emission toward IRAS~16293-2422. 
According to \citet{Rao:2009} the magnetic energy associated with the magnetic field of about 4.5~mG is comparable with the rotational energy of the system given that it is a rotating disk.
Very briefly, the rotational energy (E$\rm_{r}$) of the disk divided by the magnetic energy (E$\rm_{mag}$) in the disk is given by:
\begin{equation}
\frac{E_{r}}{E_{mag}}= \frac{\frac{1}{2}r^{2}\omega ^{2} \rho \mu_{0}}{B^{2}}
\end{equation}
where r is the radius of the disk, $\omega$ is the angular velocity of the disk, $\rho$ the average density in the disk, $\mu_{0}$ the permittivity of free space and B the magnetic field. If we use the ansatz  that B$\sim$b(n$\rm_{H{_2}})^{1/2}$, where b is a constant between 1 and 5, then we obtain the result that the rotational and magnetic energies are roughly equal for b$\sim$3. Here we have used the observed quantities of r$\sim$300~AU and $\omega$ is given by $\sim$1.7$\times$10$^{-10}$~rad~s$^{-1}$.
In this connection, \citet{Hennebelle:2009} have recently performed simulations of disk formation for which both rotation and magnetic field are present. These models show that it is feasible to maintain a magnetic pseudo-disk in the presence of rotation.

The formation of a pseudo disk and its growth are regulated by the geometry of the magnetic field \citep[see][]{Davidson:2011}.
The rotational axis of such a disk should be aligned with the magnetic field direction \citep{Galli:1993,Crutcher:2006,Davidson:2011}.  
Recently, \citet{Alves:2012} using observations of H$_{2}$O masers at 22~GHz carried out with the Very Large Array (synthesized beam of 0.14$\arcsec$ $\times$ 0.08$\arcsec$) described the magnetic field structure around IRAS16293A, assuming that the H$_{2}$O polarization vectors are parallel to the direction of the magnetic field in the plane-of-the-sky \citep{Alves:2012}. Comparison between our eSMA and SMA observations and these H$_{2}$O linear polarization vectors  shows that the configuration of our supposed pseudo-disk symmetry axis is aligned with the magnetic field direction. Our C$^{17}$O and C$^{34}$S observations are thus consistent with this scenario.

%
\section{Conclusions}
\label{sec:Conclusions}
We have performed a subarcsecond (0.59$\arcsec$ $\times$ 0.38$\arcsec$) interferometric study of the velocity structure of the low-mass protostar IRAS~16293-2422  using combined SMA and eSMA observations of  C$^{17}$O (3--2) and C$^{34}$S (7--6). Our main results and conclusions are the following:
\begin{enumerate}
\item A velocity gradient which is oriented in a northeast--southwest direction is observed towards source IRAS16293A. More specifically, this northeast-southwest velocity gradient prevails in the bulk of the C$^{17}$O and C$^{34}$S emission which is composed of blue and red-shifted emissions lying in the $v$$\rm_{LSR}$ range $-$3 to 9~km~s$^{-1}$.
\item Our observations show that the C$^{17}$O and C$^{34}$S emissions are probing larger scales than HCOOCH$_{3}$ and are therefore consistent with having a larger enclosed mass. In addition, the HCOOCH$_{3}$ ALMA-SV observations show that if the mass of the central object is greater than 0.1~M$_\odot$ then the Keplerian velocity field will be inconsistent with our measurements.
\item The C$^{17}$O and C$^{34}$S observations appear to probe a rotating structure. 
This structure and the dynamics of the gas could result from the presence of a magnetic field through formation of a magnetic pseudo-disk.
\end{enumerate}

The data presented in this paper illustrate the necessity of high angular resolution observations with high spectral resolution combined with single-dish observations (to recover the extended emission) to disentangle the motion of the gas in this object and understand which scenario prevails here.
The data also show that the structure of the low-mass protostar IRAS~16293-2422 is complicated and therefore only a complex model of the source will help us to constrain and access the relative importance of outflowing, infalling and rotational motions.

%

%
\acknowledgments

We would like to thank the entire SMA and eSMA staff who produced such excellent instruments. The development of the eSMA has been facilitated by grant 614.061.416 from the Netherlands Organisation for Scientific Research, NWO. The Submillimeter Array is a joint project between the Smithsonian Astrophysical Observatory and the Academia Sinica Institute of Astronomy and Astrophysics and is funded by the Smithsonian Institution and the Academia Sinica. We are grateful to Sandrine Bottinelli who was the original proposer of the presented eSMA observations. C.~F. thanks Edwin Bergin for enlightening discussions. C.~F. also acknowledges the financial support  provided by The Instrument Center for Danish Astrophysics (IDA). The research of J.~K.~J. was supported by a Junior Group Leader Fellowship from the Lundbeck foundation. Research at Centre for Star and Planet Formation is funded by the Danish National Research Foundation. This research used the facilities of the Canadian Astronomy Data Centre operated by the National Research Council of Canada with the support of the Canadian Space Agency. 


{\it Facilities:} \facility{SMA}, \facility{ESMA}, \facility{ALMA}, \facility{JCMT}

%
\bibliographystyle{apj}

\end{document}